\documentclass[11pt,letterpaper]{article}
\usepackage{color}
\usepackage{amsmath}    
\usepackage{amssymb}
\usepackage{graphicx} 
\DeclareGraphicsExtensions{.png,.jpg,.mps,.gif,.bmp}
\usepackage{float}
\usepackage{authblk}
\usepackage[a4paper]{geometry}
\usepackage[final]{pdfpages}

\begin{document}     

\title{Identification of epileptic regions from
electroencephalographic data: Feigenbaum graphs}

\author[]{Gabriel Guarneros B.} 
\author{Cristian P\'erez A.} 
\author{Andrea Montiel P.}
\author{J. F. Rojas}
\affil[]{Facultad de Ciencias F\'isico Matem\'aticas.\\ Benemerita Universidad Aut\'onoma de Puebla,\\Av. San Claudio y 18 sur, Ciudad Universitaria, Col. San Manuel.\\ C. P. 72570. Puebla, M\'exico.}
  

\date{}  
\maketitle  
  
\begin{abstract}

  Diagnosing epilepsy is a problem of crucial importance. So analysing EEG data is of much importance to help this diagnosis. Assembling the Feigenbaum graphs for EEG signals. And calculating their average clustering, average degree, and average  shortest path length. We manage to characterize two different data sets from each other.

Each data set consisted of focal and non-focal activity, from where epileptic regions could be identified. This method yields good results for identifying sets of data from epileptic zones. Suggesting our approach could be used to aid physicians with diagnosing epilepsy from EEG data.

\end{abstract}




\section{Introduction}
\label{S:1}

Epilepsy is a disease that affects sixty five million people in different countries \cite{ep-foun}. And two and a half million new cases are detected every year \cite{ep-foun}. Epilepsy is a disease characterized, by an enduring predisposition to generate epileptic seizures. And by the neurobiological, cognitive, psychological, and social consequences of this condition \cite{Buck1997-ys}.\\

Epileptic people are two or three times more likely to die prematurely \cite{Ridsdaleh718}. And $50\%$ of the cases, begin in childhood or adolescence \cite{PMID22972641}. 

Epilepsy is characterized by seizures, which can affect persons of any age. The seizures can be as sparse as once a year, or as often as several times a day \cite{doi:10.1111/j.1528-1157.1993.tb02586.x}. Seizure disorders are not necessarily epilepsy. Or in other words, not all seizures are epileptic fits \cite{Bodde2009-ui,pmid25944112}.\\

Epileptic seizures are unprovoked, due the involvement of the central nervous system. And, non-epileptic seizures could be due to several measurable causes \cite{Bodde2009-ui}. Such as stroke, dementia, head injury, brain infections, congenital birth defects, birth-related brain injuries, tumors and other space occupying lesions \cite{Bodde2009-ui,pmid17280251}.

Given this factors, the importance of diagnosing epilepsy is very high \cite{doi:10.1002/ana.1032}. And so, are the tools and techniques used and developed for this end \cite{DUTTA2014155,LEHNERTZ20147}.

One of the procedures for diagnosing  epilepsy, comes from analysing the EEG of a patient \cite{pmid30101347}. The EEG measures the electrical activity of the cortical area, by means of electrodes placed on the scalp of the patient \cite{pmid2262540}. More accurately, it measures the electrical potential of the dendrites of the neurons adjacent to cortical surface. Hence, the relevance of EEG analysis in diagnosing neural disorders, and epilepsy in particular \cite{GHOSH20141}.\\

Due to the fact that, EEG recordings are in essence a time series with lots of noise \cite{PhysRevE.64.061907}. The task of analysing, and achieving a diagnosis,  becomes a very difficult one \cite{FAUST201556,Liang:2010:CEC:1840667.1928477}. And requires a very well trained physician \cite{doi:10.1002/ana.1032}.

That is why many scientists, are trying to develop techniques to ease this workload, and facilitate the physician's job \cite{ACHARYA2013147,AHMADLOU2014694,pmid25997732}. 

One field of study of much relevance, is the automated EEG analysis. Which includes many computer aided algorithms, such as: component analysis \cite{noauthor_2007-rm}, Fourier Transform \cite{Welch1967-kw,pmid30084925}, wavelet transform \cite{pmid12581851,4291661}, and entropy analysis\cite{pmid26931599,entropiescad,pmid23245676} among others \cite{Welch1967-kw,Vetterli1992-fq,entropiescad,Amezquita-Sanchez:2015:NMW:2822911.2822934,pmid25997732}. 
\\

Resenty Zhong-Ke Gao et al \cite{Gao2017-dv} presented a time-frequency visibility graph to classify epileptiform EEG data. They build the Adaptive Optimal time-frequency representation of the EEG, that renders a diagnostic energy distribution. Then, they assemble the visibility graphs, and classify them. Mainly by clustering coefficient, degree and coefficient entropy. Where they manage to achieve promising results to classify between, data with epileptiform activity and data without it.\\

Guohun Zhu et al \cite{Zhu2014-ua} characterized sleep stages from EEG data. Using visibility graphs, and degree distributions for segments of single channel EEG recordings. They assembled the visibility graph, and subtracted a horizontal visibility graph edges, to obtain essential degree sequences. With  this, they manage to obtain an $87.5\%$ accuracy for a six sleep stage classification. \\

On other hand, Deng Wang et al \cite{Wang2011-ne}  manage to identify epilepsy seizure features. Using the basis-based wavelet packet entropy method, and a cross validation with the k-nearest neighbors. They manage to identify epilepsy on EEG recordings, with accuracy of about $100\%$. Via 2- ,5-, and 10-fold cross validation.\\

In this study, we use Feigenbaum graphs to analyse EEG data from \cite{Andrzejak2012-kr}. Using statistical criteria, such as average shortest path length and average clustering coefficient. We discern between signals “F” form a focal region, and “N” from the non focal region.

The data was taken from intracranial EEG, from five different epileptic patients. It was divided into two different datasets: the F set, that is the data from the focal epileptic point. The focal point was identified as the first electrode that measure the epileptic seizure. And the “N” set, that is the data from the non focal point. A non focal point is any other point that didn’t shows the epileptic seizure first \cite{Andrzejak2012-kr}.

\begin{figure}[H]
\label{fig:rjemF}
\centering
\includegraphics[width=5in]{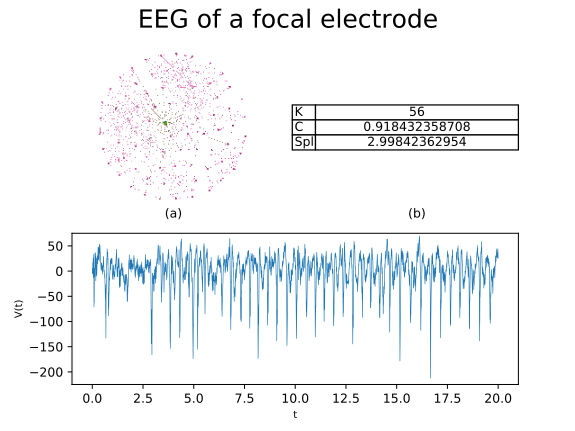}
\caption{20 seconds of EEG activity from a focal electrode. (a) shows a subnetwork for the EEG built by Feigenbaum approach, (b) shows the values for average degree $k$, average clustering $C$ and average shortest path length $spl$.}
\end{figure}

\begin{figure}[H]
\label{fig:rjemN}
\centering
\includegraphics[width=5in]{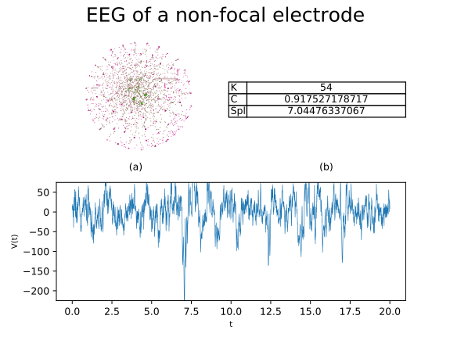}
\caption{20 seconds of EEG activity from a non-focal electrode. (a) shows a subnetwork for the EEG built by Feigenbaum approach, (b) shows the values for average degree $k$, average clustering $C$ and average shortest path length $spl$.}
\end{figure}

\section{Theory}
\label{S:2}
\subsection{Feigenbaum graphs}

The Feigenbaum graphs are a tool to characterize time series data, by constructing a network from a given time series data set \cite{Multinet,pmid24781371}. Where the structure of the network, subtracts important information from the time series \cite{Xu2008-fr,Lacasa2010-ae,Luque2011-se,GAO2012947}.

The process for building the network is as follows: for each point $x_i$ in the data set, a node $i$ is added to the network. Then for each pair of points, $x_i$ and $x_j$ in the set. Every time the criterion $x_i,x_j>x_n$, for all $n$, such that $i<n<j$ is met. An edge is added between nodes $i$ and $j$ \cite{Bodde2009-ui,Luque2011-se}.

\begin{figure}[H]
\centering
\includegraphics[width=1.0\linewidth]{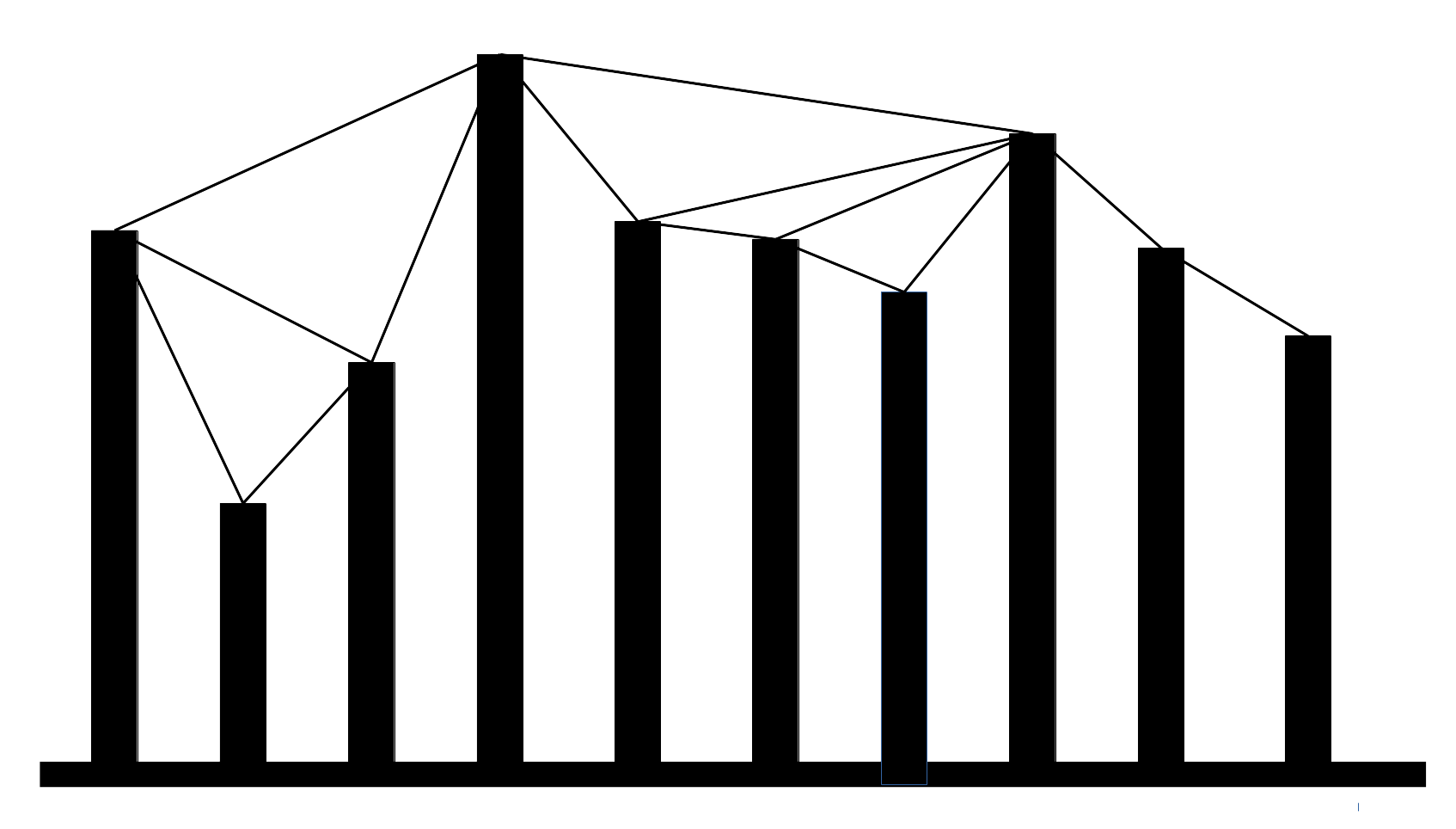}
\caption{A data set where the lines indicate a link in the network.}
\label{fig:red_ejemplo_datos}
\end{figure}

\begin{figure}[H]
\centering
\includegraphics[width=1.0\linewidth]{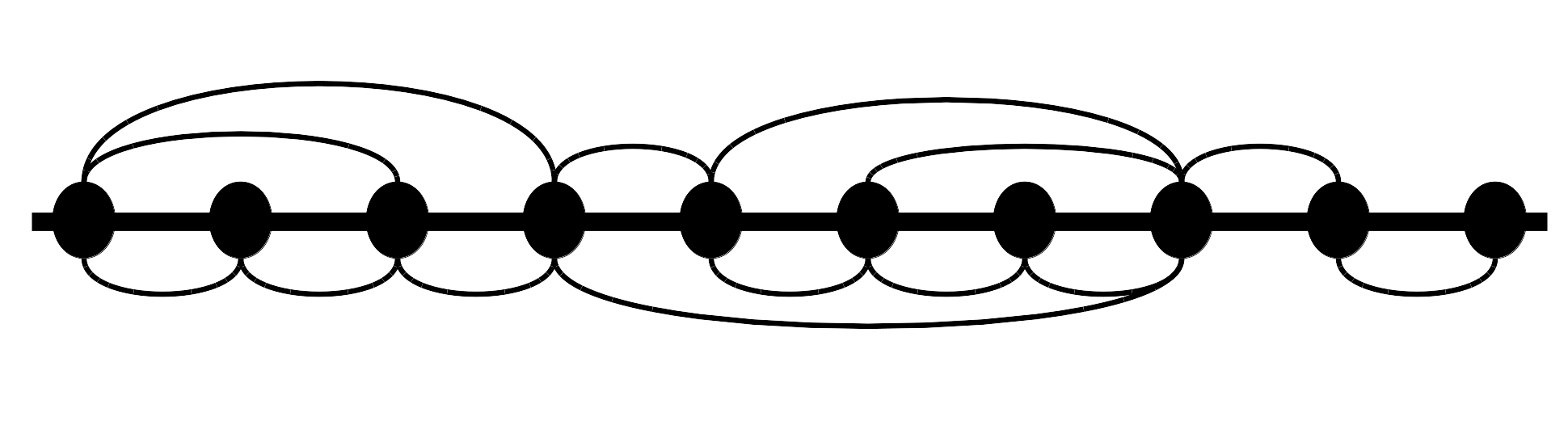}
\caption{The network that results from the data in  figure \ref{fig:red_ejemplo_datos}, by following the aforementioned procedure.}
\label{fig:red_ejemplo}
\end{figure}

\begin{figure}[H]
\centering
\includegraphics[width=1.0\linewidth]{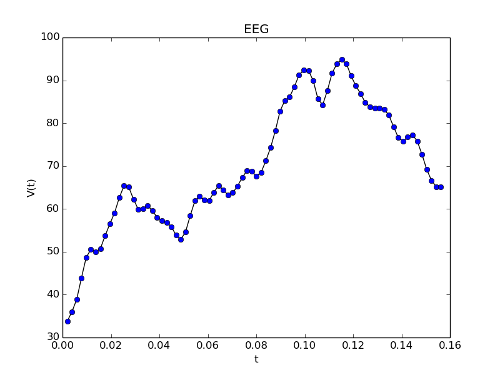}
\caption{EEG data from the \cite{Andrzejak2012-kr} data set.}
\label{fig:EEG}
\end{figure}

\begin{figure}[H]
\centering
\includegraphics[width=1.0\linewidth]{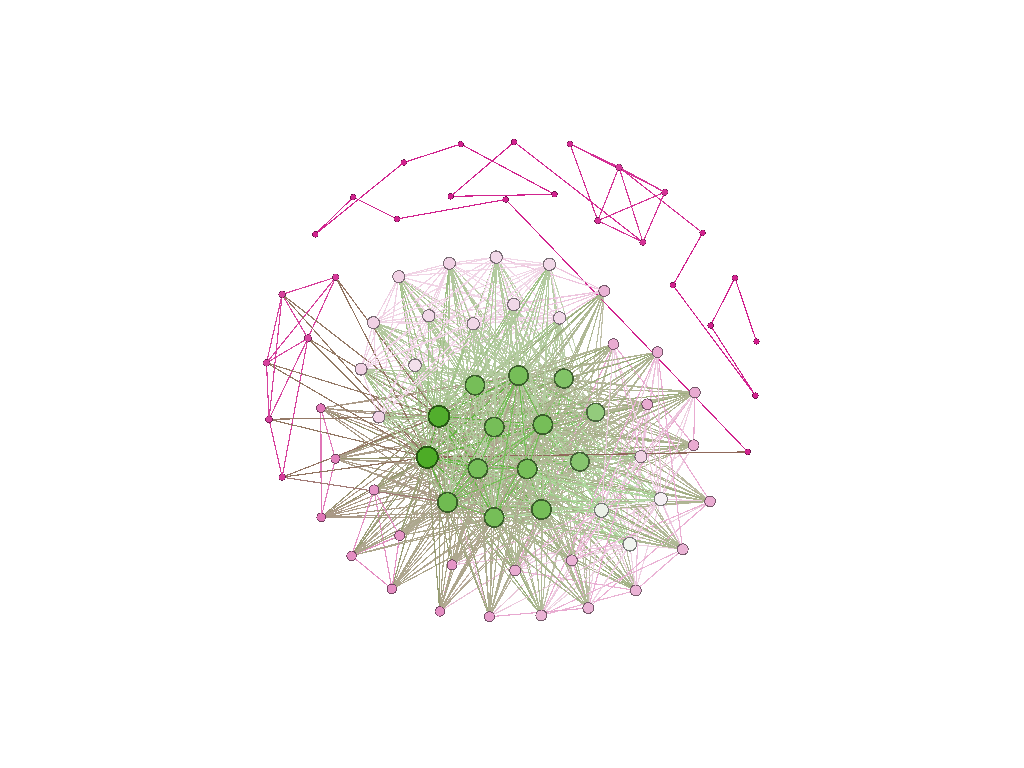}
\caption{The network that results from the data in figure \ref{fig:EEG}, by following the aforementioned procedure.}
\label{fig:red_EEG}
\end{figure}

Take the EEG from figure \ref{fig:EEG}. For each data point, a node is added to the network in figure \ref{fig:red_EEG}. And the edges are created following the procedure in figures \ref{fig:red_ejemplo_datos} and  \ref{fig:red_ejemplo} .

\section{Statistical measurements}
\label{S:3}

Having assembled the Feigenbaum graphs, we proceeded to analyse them. To find a measure to characterize them as “N” or “F” whichever was the case.

For this end, we calculated the average shortest path length for each graph. Which has a direct correlation to the size of the graph, and the data itself. On other hand, we calculated the average clustering coefficient. That is a measurement, of how the network is connected. And has correlation with how auto-similar the data is.

The average clustering coefficient is calculated by means of the equation \ref{eq:clusteringP} \cite{PhysRevE.75.027105,1367-2630-10-8-083042}.

\begin{equation}
\label{eq:clusteringP}
C=\frac{1}{n}\sum_{v \in G}{c_v}
\end{equation}

Where $n$ is the number of nodes in the network $G$, and $c_v$ is the clustering coefficient for each node $v$.

The clustering coefficient $c_v$ is calculated by means of equation \ref{eq:clustering} \cite{PhysRevE.71.065103}.

\begin{equation}
\label{eq:clustering}
c_v=\frac{2T(v)}{k(v)(k(v)-1)}
\end{equation}

Where $T(v)$ is the number of triangles through node $v$, and $k(v)$ is the degree of $v$.

Then the average shortest path length is calculated by means of the equation \ref{eq:aspl} \cite{Chen:1996:DAS:242224.242246}.

\begin{equation}
\label{eq:aspl}
a=\sum_{s, t \in V}{\frac{d(s,t)}{n(n-1)}}
\end{equation}

Where $V$ is the set of nodes in the graph, $d(s,t)$ is the shortest path length from $s$ to $t$. And $n$ is the number of nodes in the graph.

As we obtained the average shortest path length, and the average clustering coefficient for each signal in each data set. We then, assembled the distributions for the shortest path measurement in each data set. Shown in figures \ref{fig:spl_F} and \ref{fig:spl_N}.

\section{Results and Discussion}
\label{S:4}

For each EEG signal in the data set, a feigenbaum graph was generated. Once all the graphs for each signal type (focal “F” and non-focal “N”) were obtained. The average shortest path length  is calculated. The same as average clustering coefficient.

\begin{figure}[H]
\caption{Distributions of average shortest path length \ref{fig:spl_F} and average clustering coefficient \ref{fig:C_F} for focal data.}
	\centering
	\includegraphics[width=0.9\linewidth]{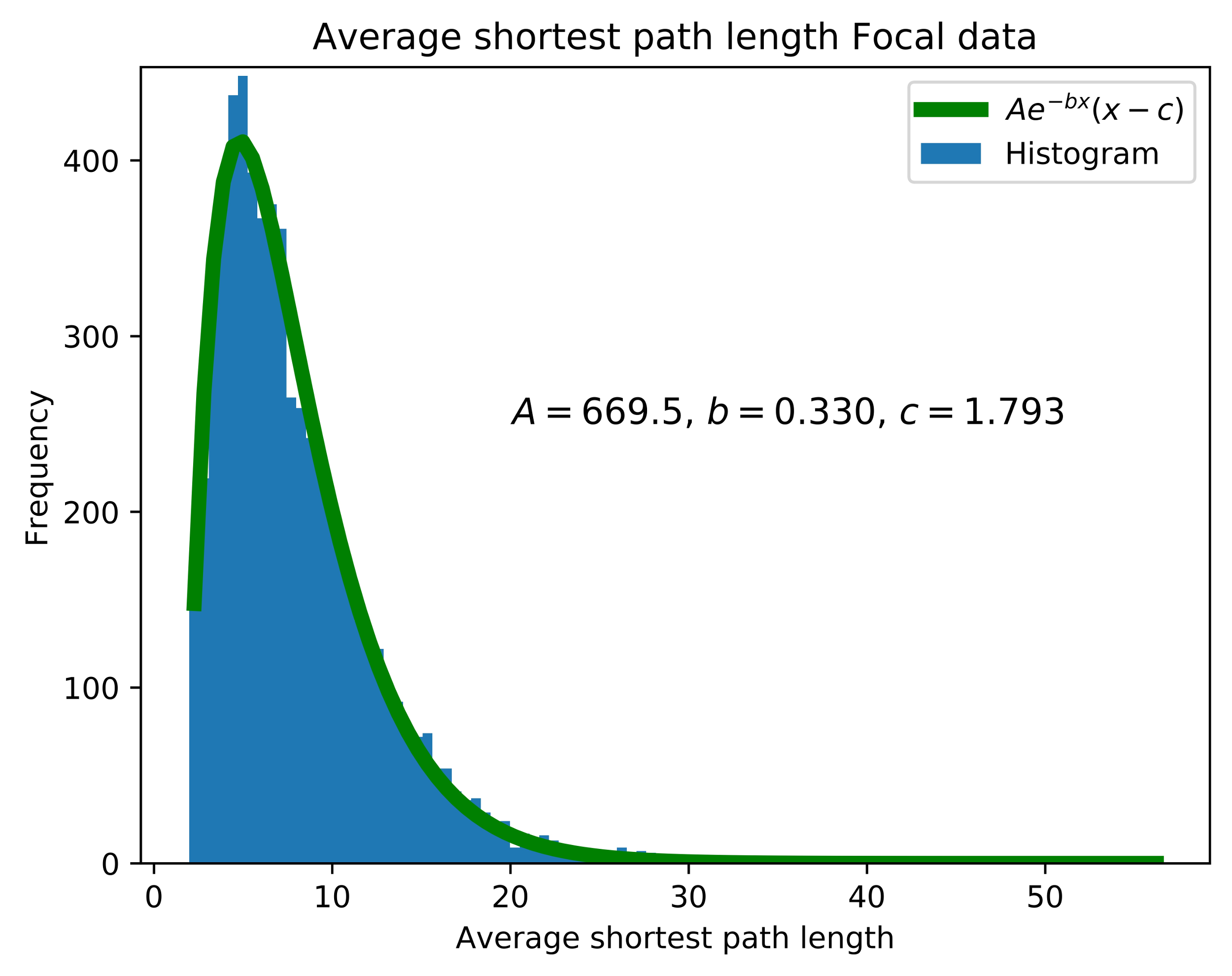}
	\caption{Distribution of the shortest path length for the focal data set. Curve fitted to $A\exp(-bx)(x-c)$ where $A=669.5$, $b=0.330$ and $c=1.793$.}
	\label{fig:spl_F}
\end{figure}

\begin{figure}
	\centering
	\includegraphics[width=0.9\linewidth]{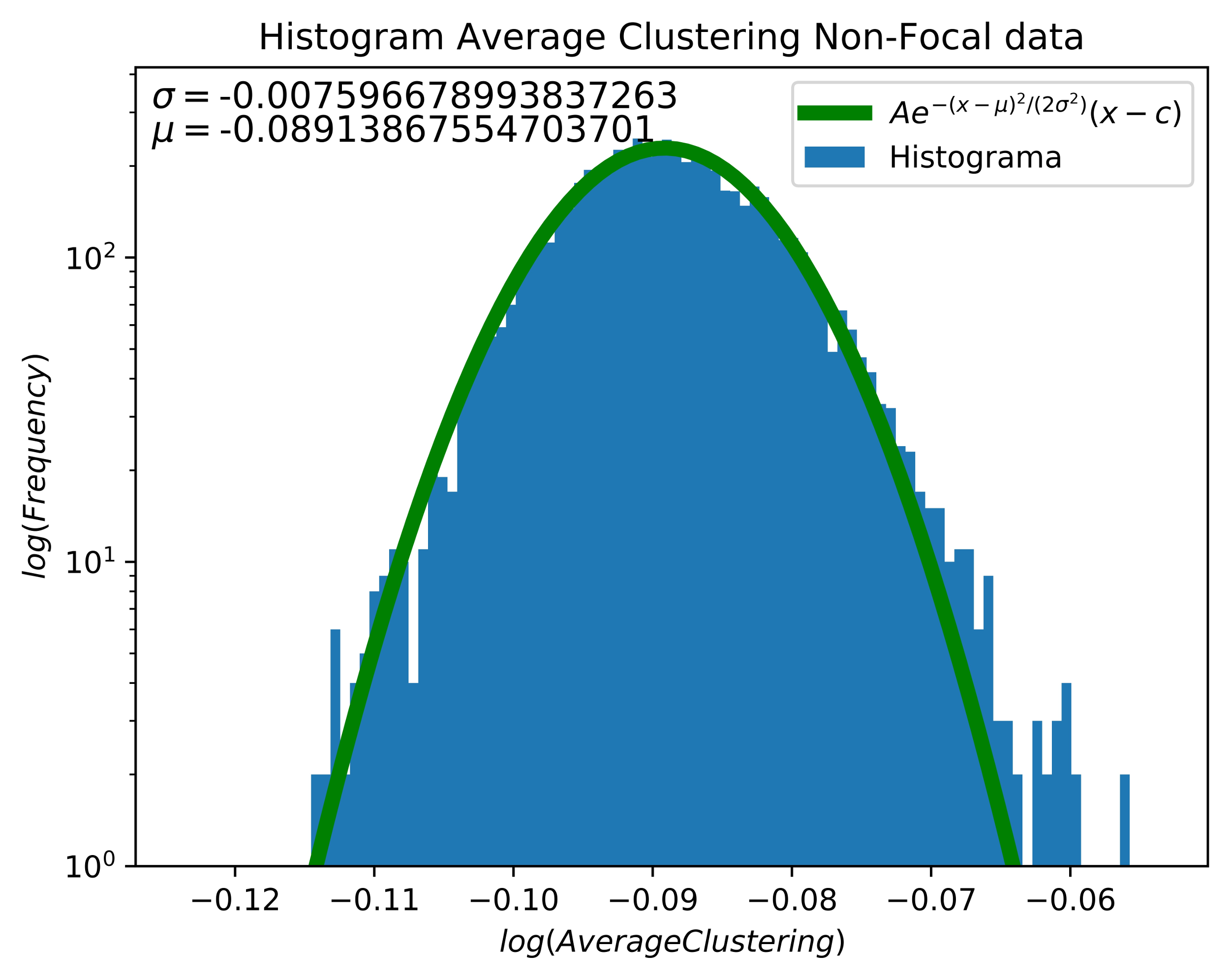}
	\caption{ Average clustering distribution for the focal data set. Curve fitted to $A\exp(-(x-\mu)^2/(2\sigma^2))$, where $\sigma=-0.007$ and $\mu=-0.089$.}
	\label{fig:C_F}
\end{figure}

\begin{figure}[H]
\caption{Distributions of average shortest path length \ref{fig:spl_N}, and average clustering coefficient \ref{fig:C_N} for non-focal data.}
	\centering
	\includegraphics[width=0.9\linewidth]{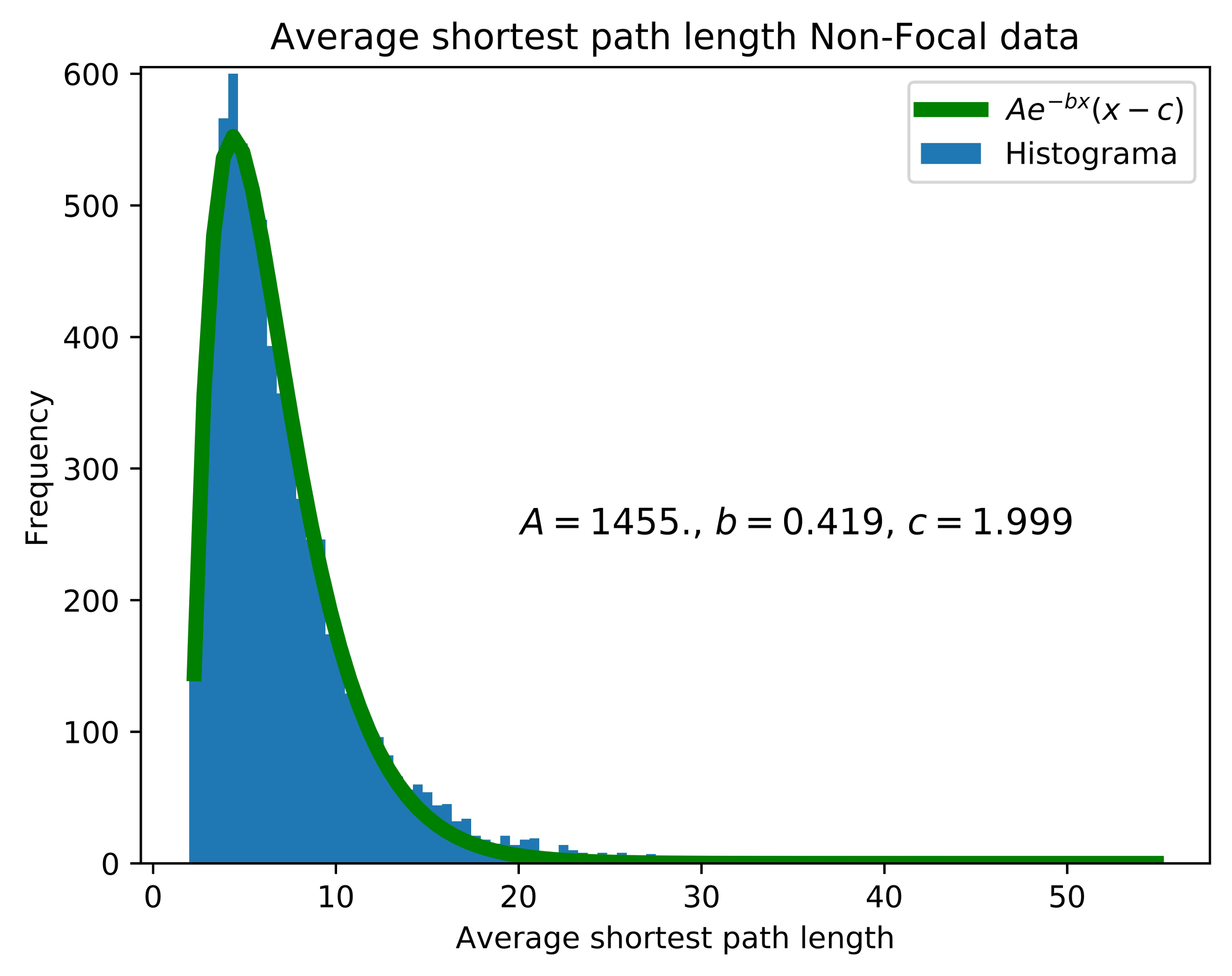}
	\caption{Distribution of the shortest path length for the non-focal data set. Curve fitted to $A\exp(-bx)(x-c)$ where $A=1455$, $b=0.419$ and $c=1.999$.}
	\label{fig:spl_N}
\end{figure}

\begin{figure}
	\centering
	\includegraphics[width=0.9\linewidth]{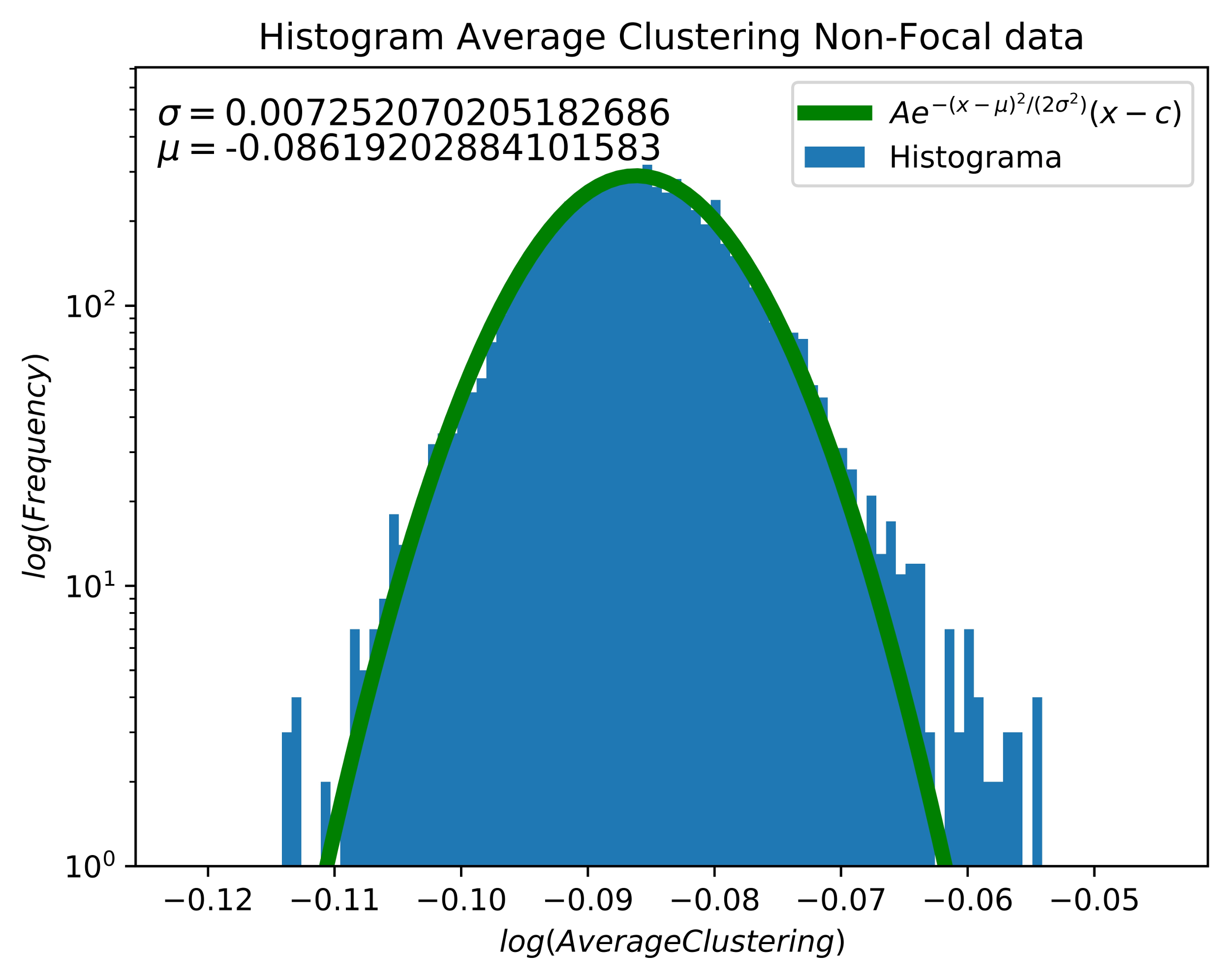}
	\caption{Average clustering distribution for the non-focal data set. Curve fitted to $A\exp(-(x-\mu)^2/(2\sigma^2))$, where $\sigma=-0.007$ and $\mu=-0.086$.}
	\label{fig:C_N}
\end{figure}

This renders two numbers to characterize each signal. Since we aim to characterize the data set as a hole. A distribution of this single parameter values comes in hand. Hence the figures \ref{fig:spl_F}, \ref{fig:C_F}, \ref{fig:spl_N}, and \ref{fig:C_N}.

    ~

    ~

Since the difference of parameters in average clustering coefficient fits, is not significant enough. The clustering coefficient, should not be considered as a good measurement for classification and desertion between the two data sets.

However, the difference in parameters for the curve fit for the average shortest path length distributions is significant. And given the data set is assembled in such a way, that does not include epileptogenic activity \cite{Andrzejak2012-kr}. It is a measurement of different states of neural configurations.

Since activity for epileptic and non epileptic neural configurations are compiled in the database. And the difference in parameters $A$ and $b$ is significant. It suggests that it could be a measurement for epilepsy, or epileptic neural configurations, for any patient.

\section{Conclusions}
\label{S:5}

In this work we propose a new method to identify epileptic focal zones from the \cite{Andrzejak2012-kr} database. Which we manage by assembling the Feigenbaum graph. And calculating the average shortest path length, and average clustering coefficient for every dataset.

The average clustering coefficient, turns to be not helpful in the desertion from different states of the data set. The average shortest path length does.

Following the idea and the way the dataset is assembled. This measurement could be calculated for a single patient. By assembling a dataset created from segments of EEG studies, no matter the time line.

Suggesting the calculation of the curve fit for the average shortest path length distribution, of Feigenbaum graphs of the dataset. Could help the physician assess a better diagnosis for the patient.  
\\
\section*{Acknowledgments}
We are grateful for the facilities provided by the Laboratorio Nacional de Supercómputo (LNS) del Sureste de México to obtain these results.

\bibliographystyle{unsrt}
\bibliography{references}

\end{document}